\documentclass[a4paper,12pt]{article}
\usepackage{graphicx}
\usepackage[T1,T2A]{fontenc}
\usepackage[utf8]{inputenc}
\usepackage{amssymb}
\usepackage{amsmath}
\usepackage{setspace}
\usepackage{xcolor}
\graphicspath{{images/}}
\begin{document} 

\thispagestyle{empty}

\begin{center}
{\large
{\bf 
Nontrivial Ground State Degeneracy of the Spin–Pseudospin Model of a Two-Dimensional Magnet Near the Frustration Point
} 
}
\vskip0.5\baselineskip{
\bf 
D.~N.~Yasinskaya$^{*1}$, 
V.~A.~Ulitko$^{1}$,
Yu.~D.~Panov$^{1}$
}
\vskip0.1\baselineskip{
$^{1}$Ural Federal University 620002, 19 Mira Street,  Ekaterinburg, Russia
}
\vskip0.1\baselineskip{
$^{*}$daria.iasinskaia@urfu.ru
}
\end{center}

The classical Monte Carlo method is used to study the properties of the ground state and phase transitions of the spin–pseudospin model, which describes a two-dimensional Ising magnet with competing
charge and spin interactions. This competition leads to the ground state degeneracy and the frustration of the
system. It is shown that the ground state degeneracy is observed in the frustration area with nonzero probabilities of the formation of two different ordered states. Based on the histogram analysis of the Monte Carlo
data, the type of phase transitions is analyzed. It is found that, near the frustration point, first-order phase
transitions are observed in the dependence on the ratio between the spin ($s = 1/2$) and pseudospin ($S = 1$)
interactions.\\


\section{Introduction}
The problem of elucidation of the microscopic
nature of unusual properties of systems with strong
interparticle interactions becomes more complex as
there is competition and/or coexistence of various
types of ordering. This problem is topical for hightemperature superconducting (HTSC) cuprates in
which the static magnetic order and the charge density
waves coexist~\cite{Fradkin2015}. The systems with various degrees of
freedom are usually described using pseudospin models~\cite{Moskvin2011,Moskvin2015}.

The spin–pseudospin model considered in this
work was proposed in~\cite{Panov2016} for studying the competition
of the charge (related to pseudospin) and spin degrees
of freedom in a quasi-two-dimensional HTSC cuprate
in the normal state. This model, like many other pseudospin models, is also suited to the description of
physical properties of dilute frustrated magnets. Due
to the flexibility in the definition of the physical
meaning of pseudospin operators, pseudospin models
are widely used for describing the properties of a great
variety of physical systems. An example is the known
Blume–Emery–Griffiths model~\cite{BEG}, which is used for
the description of properties of binary alloys~\cite{Newman1983}, multicomponent liquids and dilute magnets~\cite{Sivardiere1975}, and also
cold atoms and superconducting systems~\cite{Loois2008,Cannas2019}. Even
simple lattice models, such as the Ising or Potts models can describe a wide class of real physical systems in
the dependence on the physical meaning of pseudospins~\cite{Diep2013}.

The competition between charge (pseudospin) and
magnetic (spin) orderings in the system under study
leads to the system frustration. The systems with two
and more competing interactions, each of which leads
to a certain type of ordering, have a wide variety of
ordered phase states with various thermodynamic
properties and complex symmetry. In addition, the
frustrated systems have a high sensitivity to external
influences, fields, anisotropy, and impurities~\cite{Diep2013,Kaplan2007}
and demonstrate changes in the critical behavior~\cite{Kalz2012}.
Moreover, the models of frustrated magnets are interesting for their close connection with spin liquids,
glasses~\cite{Balents2010}, and ices~\cite{Bramwell2001}.

\section{Model and Methods}
The spin–pseudospin model~\cite{Panov2016} is a system of
mixed type with spin $s = 1/2$ and pseudospin $S = 1$. We
use pseudospin to describe valence states of the CuO$_2$
planes of HTSC La$_{2-x}$Sr$_x$CuO$_4$ cuprates in the framework of the pseudospin formalism. Here, pseudospin
$S = 1$ is a ``charge triplet'' related to three stable
valence states of [CuO$_4$]$^{5-,6-,7-}$ centers in the CuO$_2$
plane. Pseudospin projections $S_z=\pm 1$ are related to
two nonmagnetic centers [CuO$_4$]$^{5-,7-}$ with spin 0 in
the ground state. The magnetic state [CuO$_4$]$^{6-}$ is
related to pseudospin projection $S_z=0$ and is a spin
doublet with $s = 1/2$. Thus, every site of a two-dimensional lattice can be in two charge states related to
pseudospin projections $S_z=\pm 1$ and in two spin states
related to two spin projections $s_z=\pm 1/2$.

The Hamiltonian of the spin–pseudospin model of
a magnet includes the single-ion pseudospin anisotropy $\Delta$, pseudospin Ising exchange interaction $V$, and
also the conventional spin exchange interaction in the
Ising form $J$:
\begin{equation}
	\mathcal {H} = 
	\Delta \sum_i^{\phantom{N}} S_{iz}^2 
	+ V \sum_{\left\langle ij\right\rangle} S_{iz} S_{jz} 
	+ \tilde{J} \sum_{\left\langle ij\right\rangle}  \sigma_{iz} \sigma_{jz} - \mu \sum_i^{\phantom{N}} S_{iz}.
	\label{H}
\end{equation}
The summation is carried out over $N$ sites of a two-dimensional square lattice, and $\langle ij \rangle$ denotes the nearest
neighbors. Here $\sigma_{iz} = P_{0i} s_{iz}/s$ is the normalized $z$-component of spin $s = 1/2$ multiplied by projection operator $P_{0i}=1-S^2_{iz}$, which selects the magnetic [CuO$_4$]$^{6-}$
state with $S_z=0$, $\tilde{J}=Js^2=J/4$, and $\mu$ is the chemical
potential necessary to take into account the condition
of the charge constancy:
\begin{equation}
	n = \frac{1}{N} \sum_i S_{iz} = const,
	\label{constrain}
\end{equation}
where $n$ is the charge density counted from the charge
of the [CuO$_4$]$^{6-}$-center.

The ``interaction'' and the competition between
charge and spin orderings in this model are due to
kinematic restrictions related to the condition of the
completeness of the set of possible states at a given lattice site: the spin doublet with $S_{z}=0$ and $s_{z}=\pm 1/2$ and
the charge doublet with $S_{z}=\pm 1$ and $s =0$. In Hamiltonian~(\ref{H}), this correlation is taken into account in the
explicit form in projection operator $P_{0}$ entering in $\sigma_{z}$,
where it acts as the spin density operator for the
[CuO$_4$]$^{6-}$-centers.

In addition, we note that charge–charge interaction $\hat{V}$ of quite general form can be reduced to Hamiltonian~(\ref{H}) in the point charge approximation. Using
the projectors for various valence states of the CuO$_4$ center: $P_{1}$ for [CuO$_4$]$^{5-}$, $P_{0}$ for [CuO$_4$]$^{6-}$, and $P_{-1}$ for
[CuO$_4$]$^{7-}$, we write
\begin{equation*}
	\hat{V} = \sum_{\left\langle ij \right\rangle} \sum_{a,b} V_{ab} P_{ia} P_{jb} .
\end{equation*}
Assuming that for the nearest neighbors on a
square lattice $V_{ab} = V q_{a} q_{b}$, where $q_{0}$ is a charge of a center [CuO$_4$]$^{6-}$, $q_{1} = q_{0}+1$ and $q_{-1} = q_{0}-1$ are the
charges of centers [CuO$_4$]$^{5-}$ and [CuO$_4$]$^{7-}$, respectively, and taking into account the identity $S_{iz} = P_{i1} - P_{i,{-}1}$,  we obtain the expression
\begin{equation*}
	\hat{V} = V \sum_{\left\langle ij \right\rangle} S_{iz} S_{jz} + 4Vq_{0} \sum_{i} S_{iz} + 2NVq_{0}^{2} ,
\end{equation*}

which differs from~(\ref{H}) only in the shift of the energy
and chemical potential reference levels.

Thus, the system can be diluted by charged impurities which are mobile and can enter the charge–charge interaction. The properties of the system in the ground
state were studied in the mean-field approximation~\cite{MFA1}. Then, the mean-field approximation~\cite{MFA2} and the
Bethe approximation~\cite{Bethe} were used to obtain the temperature phase diagrams and the thermodynamic
characteristics. The introduction of a charge disorder
in the model can also substantially influence the critical behavior and phase states of the spin systems~\cite{YasinskayaCrit,YasinskayaPhase}. However, the aim of this work is to study the
influence of the competition of two interactions on the
formation of the ground state of an Ising magnet; thus,
we consider only the case $n = 0$.

The numerical simulation of the spin–pseudospin
system was performed using the classical Monte Carlo
(MC) method. The common Metropolis algorithm
was modified~\cite{Budrin2018} for the charge density~(\ref{constrain}) fixation to
be possible at each MC step, and also it was parallelized using the CUDA technology. The MC calculations were performed on a square lattice with periodic
boundary conditions, liner sizes $La$, the number of
sites $N = L \times L$, where $a$ is the lattice parameter taken
as 1. All calculations were carried out simultaneously
for a hundred copies of the system, to refine the results
and also to analyze the behavior of the system near the
frustration.

The temperature dependences of the specific heat
and (pseudo-)magnetic susceptibility are determined
using the fluctuation relationships
\begin{equation}
	\begin{aligned}
		C&= \left( \frac{\partial E}{\partial T} \right)_H = \frac{1}{N}\frac{\langle E^2 \rangle - \langle E \rangle^2}{k_bT^2};\\
		\chi& = \left( \frac{\partial \mathcal{O}}{\partial H} \right)_T =  \frac{1}{N} \frac{\langle \mathcal{O}^2 \rangle - \langle \mathcal{O} \rangle^2}{k_b T},
	\end{aligned}
\end{equation}
where $k_B$ is the Boltzmann constant, $E$ is the energy of
the system with Hamiltonian~(\ref{H}). Order parameter $\mathcal{O}$
for the checkerboard antiferromagnetic ($m$) and
charge-ordered ($M$) phases characterizes the spontaneous orientation ordering of the system of pseudospins and is determined as follows:
\begin{equation}
	\mathcal{O} =
	\left\{
	\begin{array}{l}
		m = m_1 - m_2,\\
		M = M_1 - M_2.\\
	\end{array}
	\right.\\
\end{equation}
Here, $m_{\lambda}=\sum\limits_{i \in \lambda} s_{iz}$ is the magnetization of sublattice $\lambda =1,2$, and $M_{\lambda}=\sum\limits_{i \in \lambda} S_{iz}$ is the mean charge
of sublattice $\lambda$ (pseudo-magnetization).

The critical temperatures of phase transitions (PT) were determined by the maxima of the specific heat and the susceptibility and also were checked using the Binder
cumulant method~\cite{Binder1992}.

\section{Results and discussion}
\begin{figure}[t]
	\center
	\includegraphics[width=\linewidth, keepaspectratio]{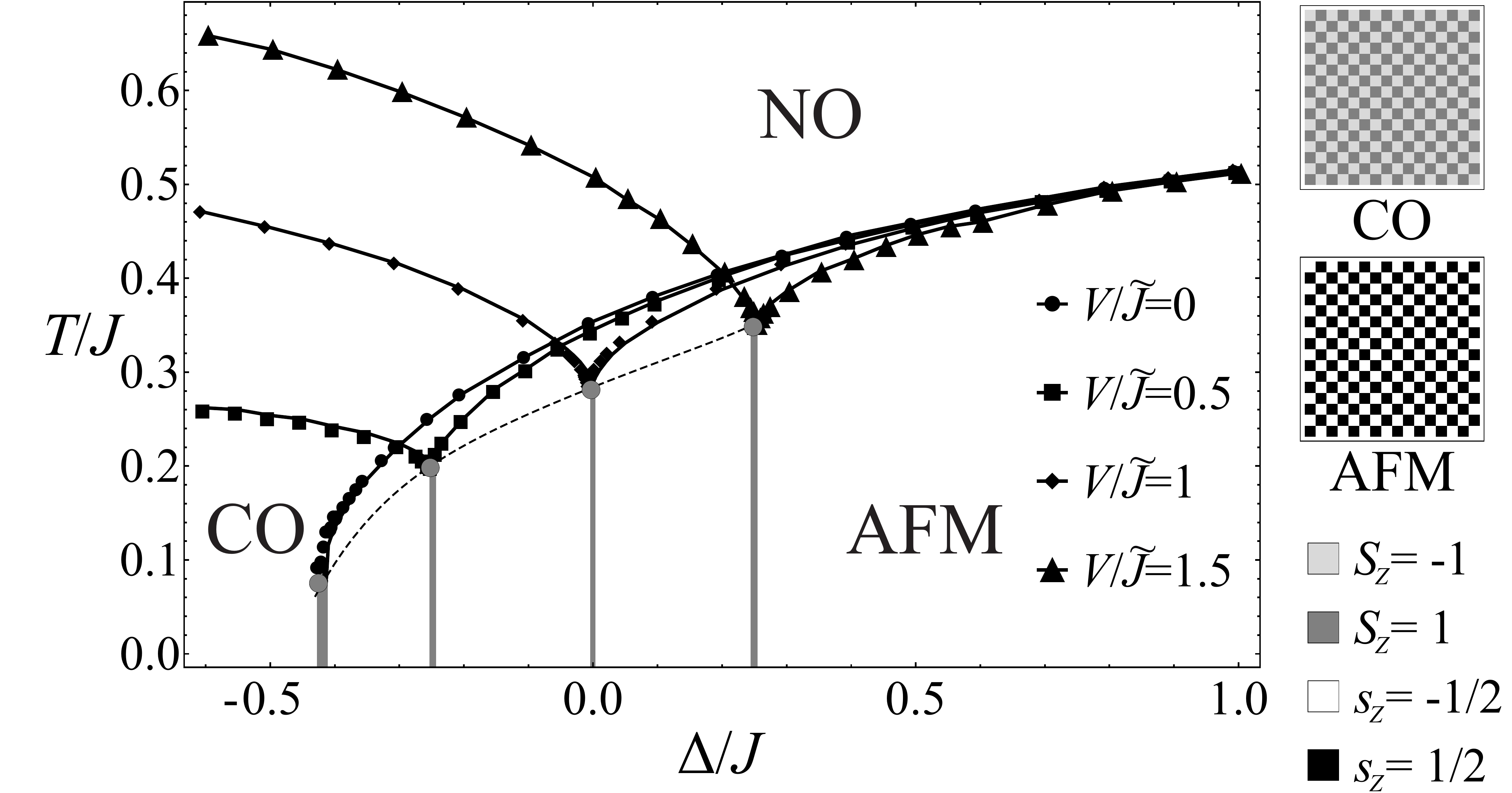}
	\caption{Temperature phase diagrams obtained in the absence of doped charge ($n = 0$) for $V/\tilde{J}=0$, $0.5$, $1$ and $1.5$. The grey points indicate the position of the frustration point $\Delta^*$. The grey vertical bands correspond to the frustration areas for every $V/\tilde{J}$ ratio}
	\label{TPDn0}
\end{figure}

Figure~\ref{TPDn0} depicts the temperature phase diagrams in
the dependence on the single-ion anisotropy parameter $\Delta/J$ for various ratios between the exchange interactions $V/\tilde{J}$. The phase diagrams have a characteristic
shape of ``checkmarks'', in the right part of which the
system is transformed from the non-ordered (NO)
paramagnetic state to the antiferromagnetic ordered
(AFM) state of spins $s = 1/2$, while, in the left part, it
is transformed to the staggered charge order (CO) of
pseudospins $S = 1$ (except for the case $V/\tilde{J}=0$, when
the charge order is impossible). At the boundary
between these two phases, the ground state of the system is degenerate because of the competition between
charge and magnetic orderings. In this system, singleion anisotropy parameter $\Delta/J$ is a frustrating one, and
the point at which the type of ordering in the ground
state is changed determines the frustration point $\Delta^*$.
This point is a classic analog of the quantum critical
point. This type of the phase diagrams of the frustrated
systems is widely distributed in the literature and, for
example, it is represented by the Ising model with the
competing interaction between the nearest and nextnearest neighbors on bcc or simple cubic lattices~\cite{Murtazaev2015,Ramazanov2018}. However, such phase diagrams were obtained for
the first time for the model with mixed spin $s = 1/2$ and
pseudospin $S = 1$.

A change in the ratio between the exchange interactions leads to a shift of the frustration point $\Delta^*$ to
larger $\Delta/J$, and this point becomes zero at $V=\tilde{J}$; the
temperature phase diagram becomes symmetric with
respect to line $\Delta^*=0$. Thus there is a boundary
between the cases of strong ($V \leq \tilde{J}$) and weak ($V>\tilde{J}$)
spin exchanges. This boundary leads to qualitative distinctions between the $\Delta/J-n$ diagrams of the ground state, and also to other effects observed as the fixed
charge density $n$ increases, as described in~\cite{MFA1,YasinskayaPhase}.

It is suggested that such phase diagrams have only
one frustration point corresponding to the minima of
the dependences of the PT temperature on the frustrating parameter. However, we revealed that the
ground state is degenerate in a whole “frustration
area”, rather than in one specific point. In this range,
both charge and antiferromagnetic orderings form in
the ground state with nonzero probabilities. This
implies that, with identical model parameters and
thermalization conditions, a part of copies of the system is ordered antiferromagnetically and another part
has the charge ordering. These areas are denoted by
grey vertical lines in Fig.~\ref{TPDn0}.

\begin{figure}[t]
	\center
	\includegraphics[width=\linewidth, keepaspectratio]{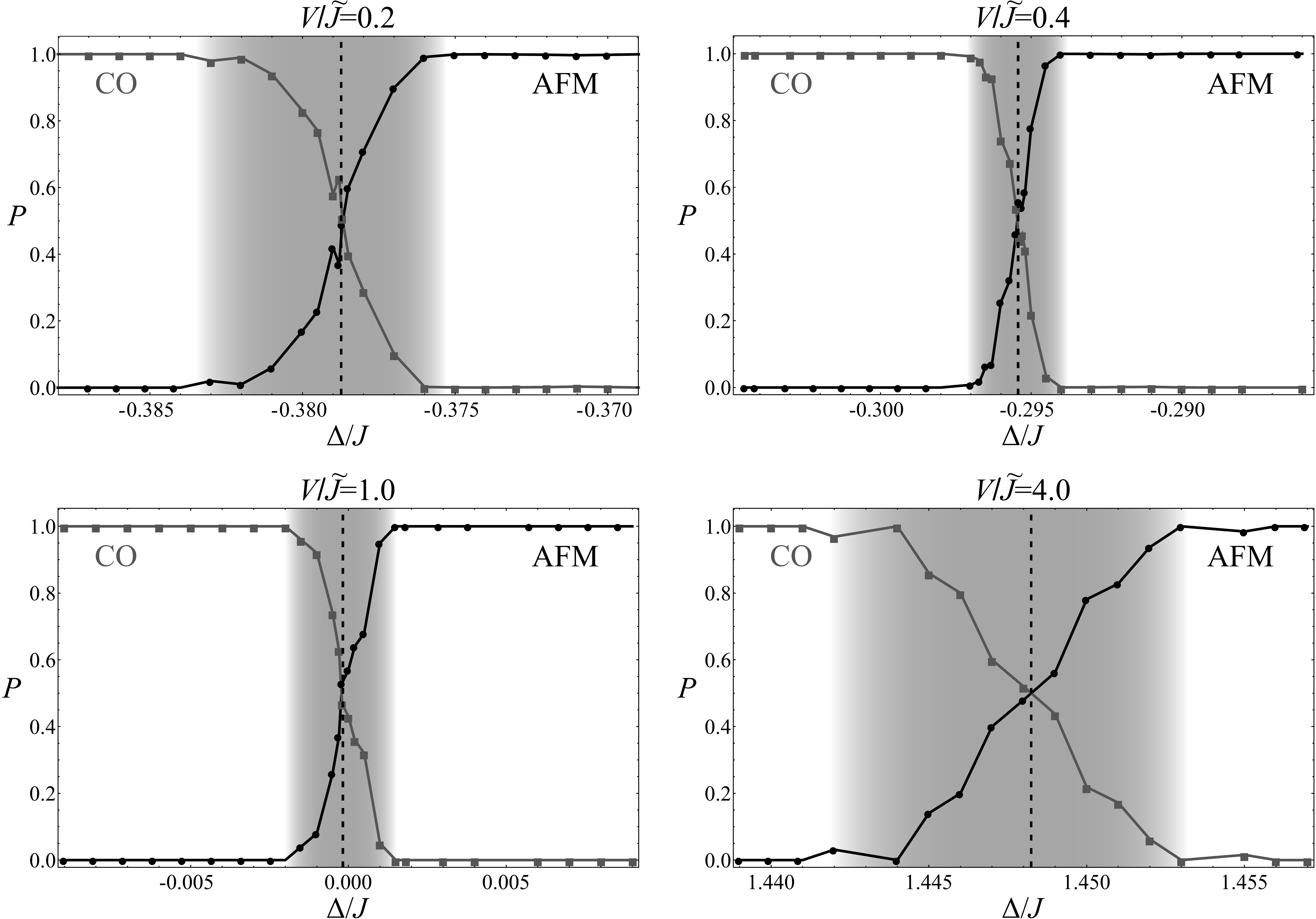}
	\caption{Distribution of the probabilities of the formation of AFM (black curve) and CO (grey curve) types of ordering in the ground state near the frustration point for $V/\tilde{J}=0.2$, $0.4$, $1$, and $4$. The grey color shows the areas in which the probability of detecting both phases in the ground state is nonzero.}
	\label{P(D)}
\end{figure}

Figure~\ref{P(D)} shows the dependences of the probabilities
of formation of the CO and AFM types of ordering for
various ratios $V/\tilde{J}$ on the frustrating parameter $\Delta/J$.
The calculations were performed for 100 copies of the
system with linear lattice sizes $L=128$ and also verified for $L=192$. Far from the frustration point, the
ground state is determined exactly: either the CO
order or the AFM order form. However, when
approaching the frustration point, the energies of
these phases become the same, and both the CO phase
and AFM phase can form in some copies of the system. Thus, the frustration point will be the value of
parameter $\Delta/J$, at which both the phases are equally
probable in the ground state. We will call the area
around the frustration point in which the probabilities
of formation of both the phases are nonzero, the frustration area (grey areas in Figs.~\ref{TPDn0} and~\ref{P(D)}).

\begin{figure}[t]
	\center
	\includegraphics[width=\linewidth, keepaspectratio]{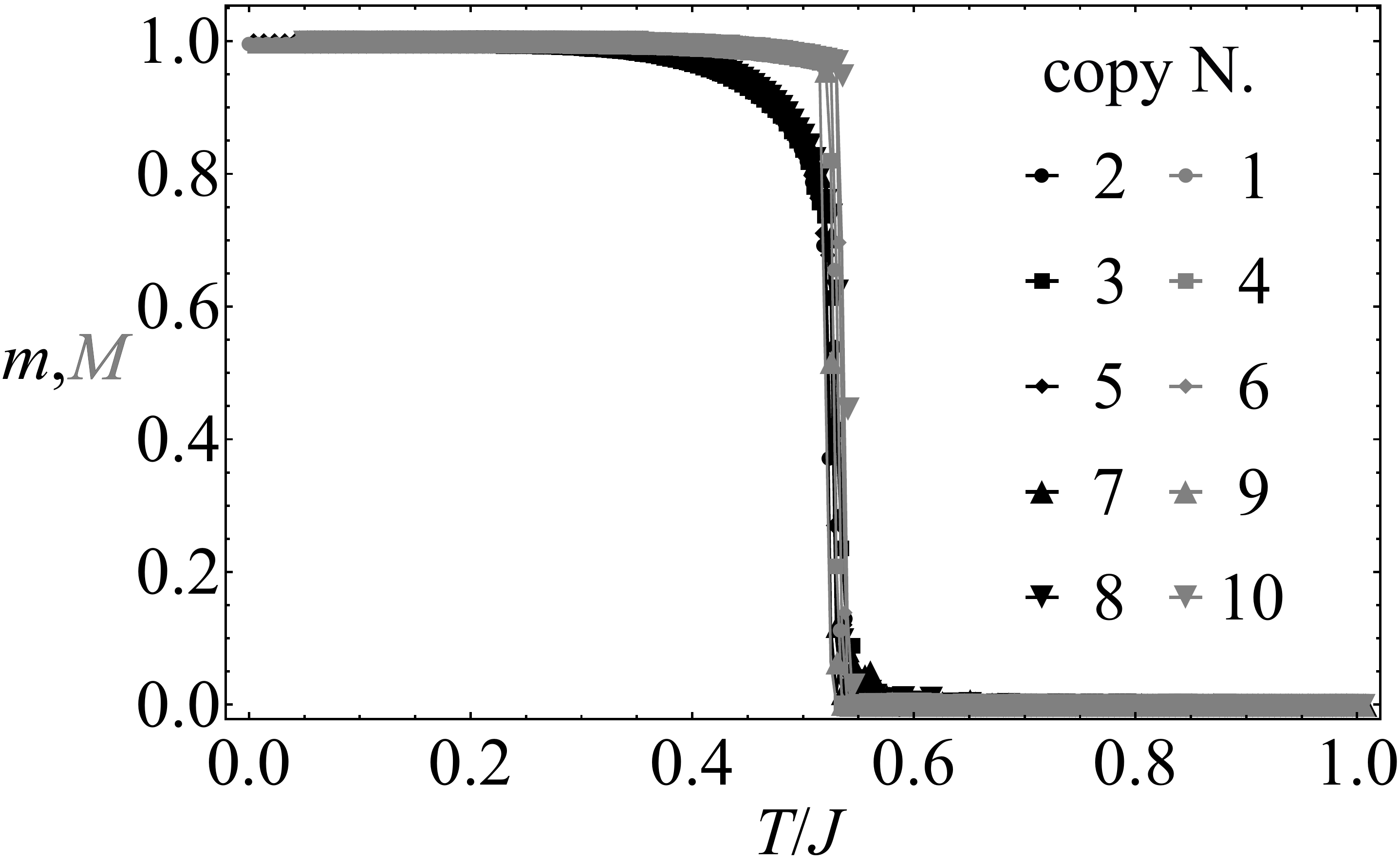}
	\caption{Temperature dependences of order parameters $m$ (black curves) and $M$ (grey curves) for ten various copies of	the system at $L=128$, $V/\tilde{J}=4$, and $\Delta/J=1.455$. A part of
		the copies transits to the AFM state, and another part transits to the CO state.}
	\label{mMfr}
\end{figure}

The competition of charge and spin interactions
leads not only to the degeneracy of the ground state
but also to the change in the type of PT. Figure~\ref{mMfr} shows
the temperature dependences of the order parameters
of AFM ($m$, black curves) and CO ($M$, grey curves)
order parameters for several various copies of the system obtained at the same parameters $L=128$, $V/\tilde{J}=4$, $\Delta/J=1.455$. In this case, the dependences of the
CO parameter $M$ behave sharper than the dependences of the AFM parameter $m$; thus, we additionally
performed the analysis of the energy histograms~\cite{WL2001,FL1991}, which allows us to determine the PT type with a
high reliability. For a first-order PT, the histogram of
energy distribution $P(E)$ near the critical temperature
$T_c$ must have a two-peak structure, each peak of which
corresponds to a certain (stable or metastable) phase,
which is demonstrated by a stepwise change in the
energy. In the case of a second-order PT, the histogram will have the form of the normal distribution,
where the most probable value of the energy corresponds to the equilibrium value of the energy of one
phase at given temperature.

\begin{figure}[t]
	\center
	\includegraphics[width=\linewidth, keepaspectratio]{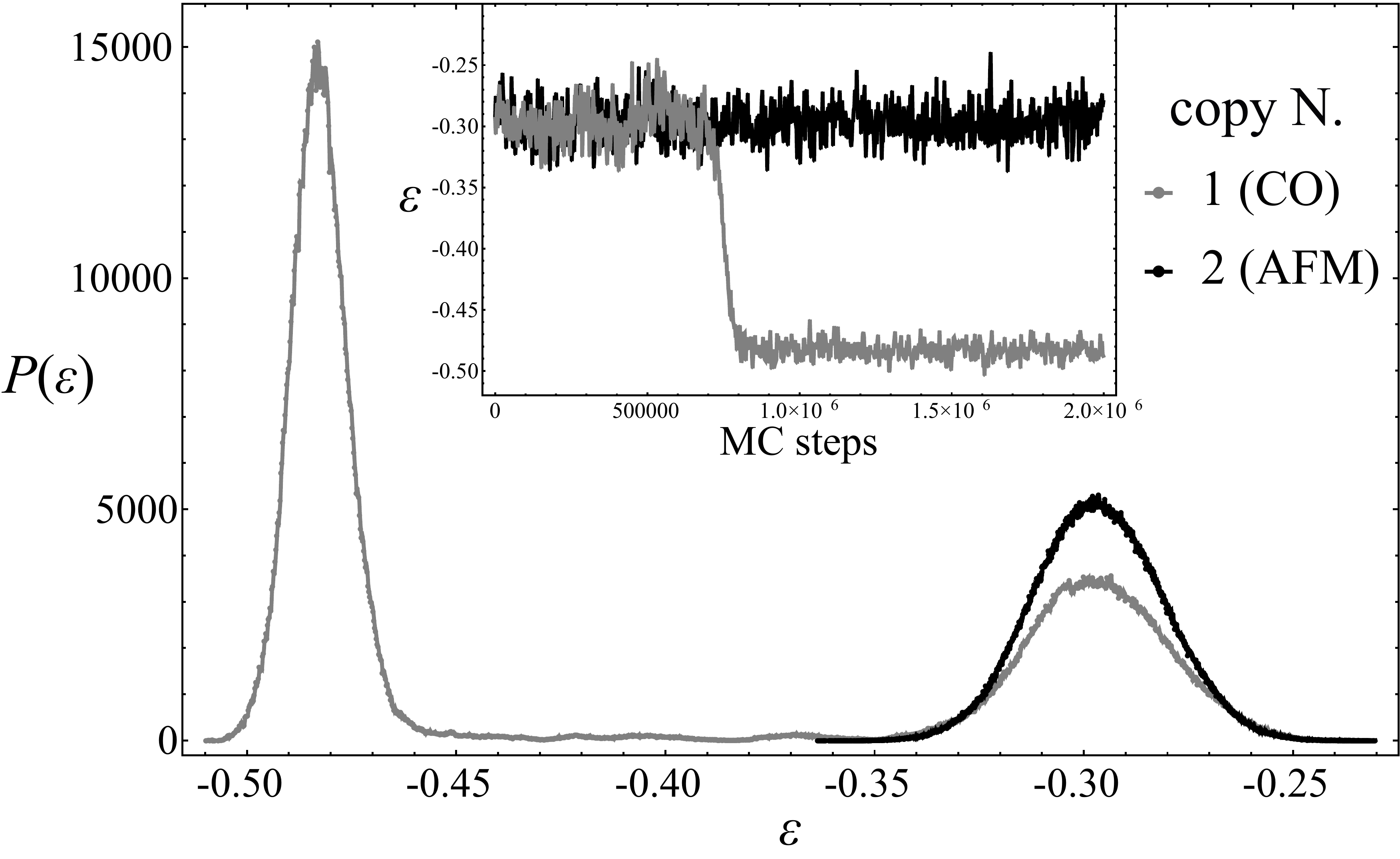}
	\caption{Histograms of the energy distribution $\varepsilon = E/(NJ)$ for two copies of the system with parameters $L=128$, $V/\tilde{J}=4$, and $\Delta/J=1.455$ at temperature $T = 0.5255J$. The
		two-peak structure of the histogram for copy 1 (grey color) demonstrates the first-order PT to the CO state. Single-peak Gaussian histogram for copy 2 (black color) shows the second-order PT to the AFM state.}
	\label{histo}
\end{figure}

Figure~\ref{histo} shows the histograms of the energy distribution $\varepsilon = E/(NJ)$ by the MC steps for copy 1 with
charge ordering and for copy 2 ordered antiferromagnetically (Fig.~\ref{mMfr}).The histograms are built near the critical point at $T=0.5255 J$. The data binning was
performed over 1000 ranges for $2 \cdot 10^6$ MC steps. The
energy distribution over MC steps is built in the inset.

The existence of two peaks in the histogram of the
energy distribution demonstrates that the transition to
the CO state in copy 1 is the first-order phase transition. The peak with a higher energy corresponds to the
non-ordered phase, and the peak with lower energy
corresponds to the charge order. One pronounced
maximum in the energy distribution histogram for
copy 2 shows that there is the second-order PT to the
AFM state.

The further analysis of the energy histograms for
various $V/\tilde{J}$ showed that, in the case of a weak spin
exchange ($V/\tilde{J}>1$), the transition to the CO state near
the frustration point is the first-order PT and the transition to the AFM state is the second-order PT. In this
case, the region width with the first-order PT becomes
larger with an increase in the $V/\tilde{J}$ ratio and the distance from point $V/\tilde{J}=1$. In the case of the strong spin
exchange $V/\tilde{J} \leq 1$, the opposite situation is observed:
the transition to the AFM state near the frustration
point is the first-order PT and the transition to the CO
phase is the second-order PT. In this case, the width
of the region with the first-order PT increases as the
distance from point $V/\tilde{J}=1$ increases and as the $V/\tilde{J}$
ratio decreases.

\section{Conclusions}
The peculiarities of the ground state and the phase
transitions of the spin–pseudospin model of a twodimensional magnet with the frustration, caused by
the competition of charge and magnetic orderings
have been studied using the classical Monte Carlo
method. It is shown that near the frustration point,
there is an area in which the probabilities of the formation both charge and antiferromagnetic phases are
nonzero. Thus, the same system can be variously
ordered at the same conditions; i.e., the ground state
of the system is degenerate in a certain frustration
area, rather than in a point.

In addition, based on the histogram analysis of the
data, it is found that the frustration influences the
types of PT. Near the frustration point in the case of
the strong spin exchange, for small $V/\tilde{J}$, the system
undergoes the first-order PT to the AFM state, and, in
the case of a weak spin exchange, for large $V/\tilde{J}$, to the
CO state.
\section{Acknowledgements}
The authors are grateful to A.S. Moskvin for stimulating discussions.

\section{Funding}
This work was supported by the Competitiveness
Enhancement Program of the Ural Federal University (Act
211 of the Government of the Russian Federation, Agreement no. 02.A03.21.0006 and CEP 3.1.1.1.g-20) and the
Ministry of Science and Higher Education of the Russian
Federation (project FEUZ-2020-0054).
%



\begin{thebibliography}{99}\setlength{\itemsep}{-0.10cm}
	
\bibitem{Fradkin2015}  E. Fradkin, S. A. Kivelson, and J. M. Tranquada, Rev. Mod. Phys. \textbf{87}, 457 (2015).

\bibitem{Moskvin2011} A. S. Moskvin, Phys. Rev. B \textbf{84}, 075116 (2011).

\bibitem{Moskvin2015} A. S. Moskvin, J. Phys.: Conf. Ser. \textbf{592}, 012076 (2015).

\bibitem{Panov2016}	Y. D. Panov, A. S. Moskvin, A. A. Chikov, and I. L. Avvakumov, J. Supercond. Nov. Magn. \textbf{29}, 1077 (2016).

\bibitem{BEG} M. Blume, V. J. Emery, and R. B. Griffiths, Phys. Rev. A \textbf{4}, 1071 (1971).

\bibitem{Newman1983} K. E. Newman and J. D. Dow, Phys. Rev. B \textbf{27}, 7495 (1983).

\bibitem{Sivardiere1975} J. Sivardi\'{e}re and J. Lajzerowicz. Phys. Rev. A \textbf{11}, 2101 (1975); Phys. Rev. A \textbf{11}, 2090 (1975).

\bibitem{Loois2008}  C. C. Loois, G. T. Barkema, and C. M. Smith, Phys. Rev. B \textbf{78}, 184519 (2008).

\bibitem{Cannas2019} S. A. Cannas and D. A. Stariolo, Phys. Rev. E \textbf{99}, 042137 (2019).

\bibitem{Diep2013} \textit{Frustrated Spin Systems}, Ed. by H. T. Diep, 2nd ed. (World Scientific, Singapore, 2013).

\bibitem{Kaplan2007} T. A. Kaplan and N. Menyuk, Philos. Mag. \textbf{87}, 3711 (2007).

\bibitem{Kalz2012} A. Kalz and A. Honecker, Phys. Rev. B \textbf{86}, 134410 (2012).

\bibitem{Balents2010}  L. Balents, Nature (London, U.K.) \textbf{464}, 199 (2010).

\bibitem{Bramwell2001} S. T. Bramwell and M. J. P. Gingras, Science (Washington, DC, U. S.) \textbf{294}, 1495 (2001).

\bibitem{MFA1} Y. D. Panov, A. S. Moskvin, A. A. Chikov, and K. S. Budrin, J. Low Temp. Phys. \textbf{187}, 646 (2017).

\bibitem{MFA2} Yu. D. Panov, V. A. Ulitko, K. S. Budrin, D. N. Yasinskaya, and A. A. Chikov, Phys. Solid State \textbf{61}, 707 (2019).

\bibitem{Bethe} Yu. D. Panov, A. S. Moskvin, V. A. Ulitko, and A. A. Chikov, Phys. Solid State \textbf{61}, 1627 (2019).

\bibitem{YasinskayaCrit} D. N. Yasinskaya, V. A. Ulitko, A. A. Chikov, and Y. D. Panov, Acta Phys. Polon. A \textbf{137}, 979 (2020).

\bibitem{YasinskayaPhase} D. N. Yasinskaya, V. A. Ulitko, and Yu. D. Panov, Phys. Solid State \textbf{62}, 1713 (2020).

\bibitem{Budrin2018}  K. S. Budrin, V. A. Ulitko, A. A. Chikov, Y. D. Panov, and A. S. Moskvin, in \textit{Proceedings of the Conference on Parallel Computational Technologies PCT’2018} (2018), p. 22.

\bibitem{Binder1992} K. Binder and D. W. Heermann, \textit{Monte Carlo Simulation in Statistical Physics} (Springer, Berlin, 1992).

\bibitem{Murtazaev2015} A. K. Murtazaev, M. K. Ramazanov, F. A. KassanOgly, and D. R. Kurbanova, J. Exp. Theor. Phys. \textbf{120}, 110 (2015).

\bibitem{Ramazanov2018} M. K. Ramazanov and A. K. Murtazaev, Phase Trans. \textbf{91}, 83 (2018).

\bibitem{WL2001} F. Wang and D. P. Landau, Phys. Rev. Lett. \textbf{86}, 2050 (2001).

\bibitem{FL1991} A. M. Ferrenberg and D. P. Landau, Phys. Rev. B \textbf{44}, 5081 (1991).
\end{thebibliography}
\end{document}